\documentclass[aps,prd,superscriptaddress,floatfix,secnumarabic]{revtex4}

\usepackage[dvips]{graphicx}
\usepackage[dvips]{color}
\usepackage{epsfig}
\usepackage{amssymb,amsmath}
\usepackage{lscape}

\newcommand{\bdk}{$B^{\pm}\to DK^{\pm}$}

\newcommand{\dkpp}{$\overline{D}{}^0\to K^0_S\pi^+\pi^-$}

\begin{document}

\title{
Feasibility study of model-independent approach to \boldmath{$\phi_3$} 
measurement using Dalitz plot analysis. 
}
\author{A. Bondar}
\affiliation{Budker Institute of Nuclear Physics, Novosibirsk}
\author{A. Poluektov}
\affiliation{Budker Institute of Nuclear Physics, Novosibirsk}
%\date{17 November, 2005}

\begin{abstract}
In this paper, we present results of a feasibility study of a model-independent
way to measure the angle $\phi_3$ of the unitarity triangle. The method 
involves \bdk\ decays where the neutral $D$ decays to the $K^0_S\pi^+\pi^-$ final state, 
together with the sample of decays of $CP$-tagged $D$ mesons (produced, 
{\it e.g. } in $\psi(3770)\to D\bar{D}$ process) to the same final state. 
We consider different approaches to the extraction of $\phi_3$
and obtain the expected statistical accuracy of the $\phi_3$ measurement as a function 
of $B$ and $D_{CP}$ statistics. 
\end{abstract}

\maketitle

\section{Introduction}

Determinations of the Cabbibo-Kobayashi-Maskawa
(CKM) \cite{ckm} matrix elements provide important checks on
the consistency of the Standard Model and ways to search
for new physics. One of the ways to check the consistency of the CKM 
model is measurement of all angles of the unitarity triangle in decays of
$B$ mesons. 
One angle, $\phi_1$, has been measured with high precision at modern 
$B$ factories by the experiments BaBar \cite{babar_phi1} and Belle
\cite{belle_phi1}. The measurement of the 
angle $\phi_2$ is more difficult due to theoretical uncertainties in 
calculation of the penguin diagram contribution. Precise determination 
of the third angle, $\phi_3$, is possible, {\it e.g. }, in the decays \bdk. 
Although it requires a lot more data than for the other angles, 
it is theoretically clean due to the absence of loop contributions. 
Finding the way to use the available experimental data most efficiently
is therefore essential for the $\phi_3$ measurement. 

Various approaches have been proposed to measure $\phi_3$ in \bdk\ decays
\cite{glw,dunietz,eilam,ads} by utilizing the interference 
between $\bar{b}\to \bar{c}u\bar{s}$ and $\bar{b}\to \bar{u}c\bar{s}$ 
transitions. The $\phi_3$ can be measured by studying the interference of 
$B^{\pm}\to D^0K^{\pm}$ and $B^{\pm}\to \overline{D}{}^0K^{\pm}$, which implies
the use of neutral $D$ final states common to $D^0$ and $\overline{D}{}^0$. 
This is provided by using $CP$ eigenstates of $D$ meson (GLW method \cite{glw})
or Cabibbo-favored and doubly Cabibbo-suppressed modes (ADS method \cite{ads}).

The Dalitz plot analysis of the three-body $D^0$ decay from \bdk\ 
processes \cite{phi3_modind,binp_dalitz,phi3_belle_1,phi3_belle_2,
phi3_belle_3,phi3_belle_4,phi3_babar_1,phi3_babar_2} 
provides today the best measurement of the 
angle $\phi_3$. It was proposed by Giri {\it et al.}
\cite{phi3_modind} and independently by the Belle collaboration \cite{binp_dalitz}. 
While Giri {\it et al.} mainly discussed a model-independent way of determining 
$\phi_3$ using a binned Dalitz analysis of \dkpp\ decay from \bdk, 
the Belle collaboration proposed an unbinned Dalitz plot fit of 
this decay using the \dkpp\ model based on a coherent sum of 
quasi-two-body amplitudes. The latter 
technique has been used so far by BaBar and Belle in their
$\phi_3$ measurement. The most recent BaBar result is 
$\phi_3=67^{\circ}\pm 28^{\circ}\mbox{(stat)}\pm 13^{\circ}\mbox{(syst)}
\pm 11^{\circ}(\mbox{model})$ using 211 fb$^{-1}$ 
data sample \cite{phi3_babar_2}, the result of the Belle measurement is 
$\phi_3=53^{\circ}\;^{+15^{\circ}}_{-18^{\circ}}
\mbox{(stat)}\pm 3^{\circ} \mbox{(syst)}\pm 9^{\circ}(\mbox{model})$
with 357 fb$^{-1}$ data sample \cite{phi3_belle_4}. 
The use of Cabibbo-favored mode 
\dkpp\ allows to increase the statistics compared to GLW or ADS methods.
However, this approach suffers from the model 
uncertainty introduced by the unknown complex phase of $D^0$ decay. 
Currently, this uncertainty is estimated to be of order of $10^{\circ}$, 
which is lower than the statistical error. In the future, 
as the statistics of $B$ decays collected at $B$-factories increases, 
this uncertainty will start to dominate the $\phi_3$ measurement. 
While the estimations of the model uncertainty can also improve, 
they can only be based on arbitrary variations of the decay amplitude 
description within phenomenological models ({\it e.g.} changing the
list of resonances or using K-matrix formalism). However these 
models are only approximations of the 3-body dynamics and thus
any model uncertainty estimates cannot be treated as absolutely reliable.

It has been shown \cite{phi3_modind} that the $D^0$ model uncertainty can be 
eliminated using the sample of neutral $D$ mesons in $CP$ eigenstates. 
This sample can be obtained at charm facilities, like currently 
running CLEO-c \cite{cleoc}, in the decays of $\psi(3770)$ into two neutral $D$
mesons. In this paper, we report the results of a feasibility study of such 
an approach using toy Monte-Carlo (MC) simulation based on the realistic 
description of $D$ decay amplitude. Our main goal is the determination 
of the amount of $D_{CP}$ decays needed for the model-independent measurement. 

The sensitivity to the angle $\phi_3$ comes
from the interference of two amplitudes producing opposite
flavors of neutral $D$ meson. Such a mixed state of neutral $D$
will be called $\tilde{D}$.
For example, in the case of $B^+$ decay, the mixed state is
$\tilde{D}_+=\overline{D}{}^0+r_Be^{i\theta_+}D^0$. Here $r_B$ is the ratio of the
suppressed and favored amplitudes,
the total phase between the $\overline{D}{}^0$ and $D^0$ is $\theta_+=\phi_3+\delta_B$,
where $\delta_B$ is the strong
phase difference between suppressed and favored $B$ decays.
Analogously, for the decay of $B^-$, one can write
$\tilde{D}_-=D^0+r_Be^{i\theta_-}\overline{D}{}^0$ with $\theta_-=-\phi_3+\delta_B$.
The ratio $r_B$ is expected to be of order of 0.1--0.2 \cite{gronau}. 
The BaBar ($r_B=0.12\pm 0.08\pm 0.03(\mbox{syst})\pm 0.04(\mbox{model})$, 
\cite{phi3_babar_1}) 
and Belle ($r_B=0.16\pm 0.05\pm 0.01(\mbox{syst})\pm 0.05(\mbox{model})$, 
\cite{phi3_belle_4}) measurements of the $r_B$ in Dalitz analysis of $D^0$
from \bdk\ decay confirm these estimates. 

The Dalitz plot density of $\tilde{D}$ gives immediate information
about $r_B$ and $\theta_\pm$, once the amplitude of the $\overline{D}{}^0$ decay is known.
The amplitude of the $\tilde{D}_{\pm}$ decay
as a function of Dalitz plot variables $m^2_+=m^2_{K^0_S\pi^+}$ and
$m^2_-=m^2_{K^0_S\pi^-}$ is
\begin{equation}
  f_{B^{\pm}}=f_D(m^2_{\pm}, m^2_{\mp})+r_Be^{i\theta_{\pm}}
              f_D(m^2_{\mp}, m^2_{\pm}),
  \label{b_ampl}
\end{equation}
where $f_D(m^2_+, m^2_-)$ is an amplitude of the \dkpp\ decay.
Here, we neglect possible effects of 
charm mixing and $CP$ violation in $D$ decay. 
It has been shown \cite{dmix}, that the mixing effects, which are 
suppressed in the Standard Model, lead to a bias in $\phi_3$ only at 
a second order in the small parameters $\Delta m_D/\Gamma_D$ and 
$\Delta\Gamma_D/\Gamma_D$  (where $\Delta m_D$ and $\Delta\Gamma_D$
are the mass and decay width differences between the two $D$ mass
eigenstates, $\Gamma_D$ is the average decay width of mass eigenstates), 
and therefore can be safely neglected. 

The \dkpp\ decay model can be determined
from a large sample of flavor-tagged \dkpp\ decays
produced in continuum $e^+e^-$ annihilation. Once that is known,
a simultaneous fit of $B^+$ and $B^-$ data allows to separate the
contributions of $r_B$, $\phi_3$ and $\delta_B$. However, flavor-tagged \dkpp\ 
decays only provide information about the absolute value of the $f_D$ 
amplitude. The phase term can only be obtained based on some model assumptions 
(such as Breit-Wigner amplitude dependence of the 
underlying resonances) which results in the model uncertainty 
of the $\phi_3$ measurement. 

In the original publication by Giri {\it et al.} \cite{phi3_modind}, 
an approach based on the binned Dalitz plot analysis was proposed
which allows to extract $\phi_3$ without model uncertainties. 
The idea of the approach is based on the fact that, 
to measure $\phi_3$, complete knowledge of the phase term of $f_D$ 
amplitude is not necessary. In the \dkpp\ decay amplitude, we emphasize 
the absolute value $|f|$ and phase $\delta_D$:
\begin{equation}
  f_D(m^2_+, m^2_-) = |f_D(m^2_+,m^2_-)|e^{i\delta_D(m^2_+, m^2_-)}
\end{equation}
Since the phase of $f_B$ is arbitrary, it can be expressed as
\begin{equation}
  f_{B^{\pm}}=|f_D(m^2_{\pm}, m^2_{\mp})|+r_Be^{\pm i\phi_3+i\delta_B}|f_D(m^2_{\mp}, m^2_{\pm})|
     \exp(i\delta_D(m^2_{\pm}, m^2_{\mp})-i\delta_D(m^2_{\mp}, m^2_{\pm})),
\end{equation}
{\it i.e.} only the difference of phase terms 
$\delta_D(m^2_+, m^2_-)-\delta_D(m^2_-, m^2_+)$ between
symmetric points of the phase space is relevant. 

This difference can be obtained either by studying the 
decays of neutral $D$ in $CP$ eigenstate or 
treating them as free parameters in the extraction of the $\phi_3$ from the 
$B$ decays (see Section~\ref{section_no_dcp}).
Decays of $D$ mesons in $CP$ eigenstate to $K^0_S\pi^+\pi^-$ can be obtained 
at $e^+e^-$ machine in the process, {\it e.g.} $e^+e^-\to\psi(3770)\to D\bar{D}$, 
where the other (tag-side) $D$ meson is reconstructed in $CP$ eigenstate, 
such as $K^+K^-$ or $K^0_S\omega$. 
The binned Dalitz plot analysis allows to determine $\phi_3$ without the 
need to parametrize the dynamics of the \dkpp\ decay ({\it e.g.} by 
introducing resonances) and thus allows for a truly model independent 
measurement. 

Different approaches to the extraction of $\delta_D$ phase and different 
Dalitz plot binnings are tested in our feasibility study and 
compared to the model-dependend approach performed in an unbinned fashion.
We consider only the approaches described in \cite{phi3_modind}. 
However, another possibility exists to access the phase information, which 
involves the process $e^+e^-\to\psi(3770)\to D\bar{D}$ with 
both neutral $D$ mesons decaying to $K^0_S\pi^+\pi^-$ final state \cite{kppkpp}. 
This approach requires to study the correlations of the two Dalitz plots.

In the next section, we follow the technique proposed in \cite{phi3_modind}, 
to introduce the notation which will be used in the rest of the paper. 

\section{Basic idea of the model-independent technique}

%The amplitude of $D^0-\overline{D}{}^0$ mixture from $B^{\pm}$ decay is given 
%by 
%\begin{equation}
%  f_{B^{\pm}}(m^2_+, m^2_-) = f_D(m^2_{\pm}, m^2_{\mp})+r_Be^{i\theta_{\pm}}f_D(m^2_{\mp}, m^2_{\pm}), 
%\end{equation}
%where $\theta_-=-\phi_3+\delta_B$ for $B^-$ decay and $\theta_+=\phi_3+\delta_B$
%for $B^+$ decay. 

The amplitude of \dkpp\ decay from $CP$ eigenstate of neutral $D$ is
\begin{equation}
  f_{CP\pm}(m^2_+, m^2_-) = \frac{1}{\sqrt{2}}[f_D(m^2_+, m^2_-)\pm 
                            f_D(m^2_-, m^2_+)], 
\end{equation}
for $CP$-even and $CP$-odd states of neutral $D$, respectively. 

Consider the \dkpp\ Dalitz plot is divided into $2\mathcal{N}$ bins 
symmetrically over exchange $m^2_+ \leftrightarrow m^2_-$. The bins will be 
denoted
by the index $i$ ranging from $-\mathcal{N}$ to $\mathcal{N}$ (excluding 0); 
the exchange $m^2_+ \leftrightarrow m^2_-$ corresponds to the exchange 
$i\leftrightarrow -i$. 
The number of events in the $i$-th bin of the Dalitz plot, given by the 
region $\mathcal{D}_i$, is 
\begin{equation}
  K_i = A_D\int_{\mathcal{D}_i}|f_D(m^2_+, m^2_-)|^2dm^2_- dm^2_+
  \equiv A_D F_i, 
  \label{dnum}
\end{equation}
where $A_D$ is the normalization that depends on the luminosity integral
and definition of amplitude $f_D$. 

Similarly, as follows from Eq.~\ref{b_ampl}, for \dkpp\ decays from \bdk\ 
mode the number of events in $i$-th bin is
\begin{equation}
\begin{split}
  N^{\pm}_i &= A_B\int_{\mathcal{D}_i}|f_{B^{\pm}}(m^2_+, m^2_-)|^2dm^2_- dm^2_+ \\
      &= A_B\left[\int_{\mathcal{D}_i}|f_D(m^2_{\pm}, m^2_{\mp})|^2dm^2_- dm^2_+ + 
      r_B^2 \int_{\mathcal{D}_i}|f_D(m^2_{\mp}, m^2_{\pm})|^2dm^2_- dm^2_+ \right. \\
      &\quad \left. + 
      2r_B Re\left(e^{i\theta_{\pm}}\int_{\mathcal{D}_i}f_D(m^2_+, m^2_-)f^*_D(m^2_-,
      m^2_+)dm^2_- dm^2_+\right)\right]\\
      &\equiv A_B\left[F_{\pm i}+2r_B Re(e^{i\theta_{\pm}} G_i)+ r_B^2 F_{\mp i}\right]\\
      &= A_B\left[F_{\pm i}+2r_B(\cos\theta_{\pm}\;Re\;G_{i} -
      \sin\theta_{\pm}\;Im\;G_{i}) + r_B^2F_{\mp i}
      \right].
\end{split}
\end{equation}
Thus in order to extract information about $\theta_{\pm}$, one needs
the values of $F_{\pm i}$, which are obtained from the decays of flavor 
tagged $D$ mesons, and also the values of $Re\;G_{i}$ and $Im\;G_{i}$.

The number of events in $i$-th bin for $CP$ eigenstate of neutral $D$ is
\begin{equation}
\begin{split}
  M^{\pm}_i &= A_{CP\pm}\int_{\mathcal{D}_i}|f_{CP\pm}(m^2_+, m^2_-)|^2 
                dm^2_- dm^2_+\\
      &= 2A_{CP\pm}(F_{i}\pm 2\;Re\;G_{i}+ F_{-i}).
\end{split}
\end{equation}

Substituting the integrals $F_i$ from Eq.~\ref{dnum} and introducing the 
parameters $c_i$ and $s_i$ which depend on the phase term $\phi(m^2_+, m^2_-)$, 
we finally obtain the following system of equations which relates the number of 
events in bins of flavor $D$, $D_{CP}$ and $D$ from $B$ decays:
\begin{equation}
  N^{\pm}_i = h_B[K_{\pm i} + r_B^2 K_{\mp i} + 2r_B\sqrt{K_{i}
  K_{-i}}(c_i\cos(\pm\phi_3+\delta_B)-s_i\sin(\pm\phi_3+\delta_B))], 
  \label{bnum}
\end{equation}
\begin{equation}
  M^{\pm}_i = h_{CP\pm}[K_{i} + K_{-i} \pm 2\sqrt{K_{i} K_{-i}}c_i], 
  \label{cpnum}
\end{equation}
where $h_B$ and $h_{CP\pm}$ are the normalization factors which take into account
different branching ratios, luminosity integrals and detection efficiencies
for $B$, $D^0$ and $D_{CP\pm}$ decays, 
\begin{equation}
  c_i = \frac{1}{\sqrt{F_{i} F_{-i}}}
        \int_{\mathcal{D}_i}|f_D(m^2_+, m^2_-)||f_D(m^2_-,m^2_+)|
        \cos(\delta_D(m^2_+, m^2_-)-\delta_D(m^2_-, m^2_+))dm^2_- dm^2_+ , 
  \label{cform}
\end{equation}
and 
\begin{equation}
  s_i = \frac{1}{\sqrt{F_{i} F_{-i}}}
        \int_{\mathcal{D}_i}|f_D(m^2_+, m^2_-)||f_D(m^2_-,m^2_+)|
        \sin(\delta_D(m^2_+, m^2_-)-\delta_D(m^2_-, m^2_+))dm^2_- dm^2_+. 
  \label{sform}
\end{equation}
Note that $c_i=c_{-i}$ and $s_i=-s_{-i}$. 
The meaning of these parameters is sine and cosine of the phase difference, 
averaged over the bin area with the weight proportional to 
$|f_D(m^2_+, m^2_-)||f_D(m^2_-,m^2_+)|$. 
It is clear that $s_i^2+c_i^2\leq 1$. 

The parameters $c_i$ and $s_i$ are in general unknown, they can be determined 
either from $D_{CP}$ decays, or by leaving them as free parameters 
in the fit to $D$ Dalitz plot from \bdk\ decay. Depending on the way 
these parameters are obtained, several approaches of $\phi_3$ extraction
are possible. Below we discuss these approaches, together with the toy 
MC simulation results for each of them. 

In our toy MC studies, we assume that the statistics of flavor-tagged 
$D$ mesons is practically infinite, therefore, the absolute value 
of the $f_D$ amplitude is known to high precision. This is provided by 
the large cross-section of $D^{*\pm}$ meson production in 
$e^+ e^-\to c\bar{c}$ process. We also assume that the number of 
$D_{CP}$ decays is much larger than the number of $B$ decays, due to the 
fact that the cross-section of $e^+e^-\to \psi(3770)\to D^0 D^0_{CP}$
production is approximately two orders of magnitude larger than 
that of $e^+e^-\to \Upsilon(4S)\to B B_{DK}$ process. 

\section{Approach with tagged \boldmath{$D_{CP}$} and constraint on $s_i$}

\label{tagged_constr_section}

It is clear that for the most precise measurement of $\phi_3$ given the 
limited statistics of $B$ decays, all the available information has to be
used. In particular, the unknown quantities $c_i$ and $s_i$ have to 
be extracted from the higher-statistics $D_{CP}$ data.

The coefficient $c_i$ can be obtained directly from the $D^0$ and 
$D_{CP}$ decays. For example, if the numbers of $D_{CP+}$ 
and $D_{CP-}$ decays are proportional to the branching fractions of 
$\psi(3770)\to DD_{CP+}$ and $\psi(3770)\to DD_{CP-}$, respectively, 
{\it i.e.} if $h_{CP+}=h_{CP-}$, 
\begin{equation}
  c_i=\frac{1}{2}\frac{M^+_i-M^-_i}{M^+_i+M^-_i}
      \frac{K_i+K_{-i}}{\sqrt{K_i K_{-i}}} 
\end{equation}
(see Eq.~\ref{cpnum}). However, $c_i$ can be extracted even in $D$ 
decays of only one $CP$ parity.

If the binning is fine enough so that both $f(m^2_+, m^2_-)$ and 
$f(m^2_-, m^2_+)$ are approximately constant over 
$(m^2_+, m^2_-)\in\mathcal{D}_i$, the expressions for 
$c_i$ (\ref{cform}) and $s_i$ (\ref{sform}) reduce to 
$c_i=\cos(\delta_D(m^2_+, m^2_-)-\delta_D(m^2_-, m^2_+))$, 
$s_i=\sin(\delta_D(m^2_+, m^2_-)-\delta_D(m^2_-, m^2_+))$. 
In such a case, $s_i$ can be directly obtained 
as $s_i=\pm\sqrt{1-c^2_i}$. The sign of $s_i$ can be chosen either 
based on the continuity requirement (it should vary smoothly 
for adjacent bins) or under the assumption that the $D^0$ decay proceeds via
two-body amplitudes. For simplicity, in our study we take the ``correct"
$s_i$ sign (calculated from the input $D^0$ model). 

Once the coefficients $c_i$ and $s_i$ are determined, the 
parameters $r_B$, $\phi_3$ and $\delta_B$ can be obtained from $B$ decays 
by minimizing the negative logarithmic likelihood
\begin{equation}
-2\log\mathcal{L}=-2\log\mathcal{L_+}-2\log\mathcal{L_-}, 
  \label{lhsum}
\end{equation}
where
\begin{equation}
  -2\log\mathcal{L_{\pm}}=\sum\limits^{\mathcal{N}}_{i=-\mathcal{N}}
    \frac{(N^{\pm}_i-h_B[K_{\pm i} + r_B^2 K_{\mp i} + 2r_B\sqrt{K_{i}
    K_{-i}}(c_i\cos(\pm\phi_3+\delta_B)-s_i\sin(\pm\phi_3+\delta_B))])^2}{N^{\pm}_i}
  \label{lh}
\end{equation}
is the negative logarithmic likelihood for one $B$ flavor ($B^+$ or $B^-$), 
$N^{\pm}_i$ is the number of events in $i$-th bin for corresponding $B$ flavor. 
Here, Gaussian behavior of the number of events in each bin is assumed. 

To avoid possible biases due to non-constant $f(m^2_+, m^2_-)$, 
the bin size should be relatively small. The size of the bin is 
determined by the width of the most narrow structure in the Dalitz plot; 
in our case, it is the $\omega$ meson with $\Gamma=8.4$~MeV$/c^2$; therefore, 
typical scale of phase variation is $\Delta m^2=0.013$ GeV$^2/c^4$. 
However, since we assume that 
the number of $B$ decays is much smaller than the number of 
$D_{CP}$ decays, the number of $B$ events per bin for such a fine binning 
can be small. This can affect the fit procedure since we expect 
Gaussian behavior of the number of events. Therefore, for our study 
we take relatively rough binning for the fit of $B$ decays, and 
introduce a sub-binning of each of "$B$-bins" into several "$D$-bins". 
As follows from definitions (\ref{cform}), (\ref{sform}) and 
(\ref{dnum}), the parameters $c_i$ and $s_i$ for "$B$-bins" can be calculated 
from finer "$D$-bins" as
\begin{equation}
  c_i=\sum\limits_{\alpha}c_{i,\alpha}\sqrt{K_{i,\alpha}K_{-i,-\alpha}}
      /\sqrt{K_{i}K_{-i}}, 
\end{equation}
\begin{equation}
  s_i=\sum\limits_{\alpha}s_{i,\alpha}\sqrt{K_{i,\alpha}K_{-i,-\alpha}}
      /\sqrt{K_{i}K_{-i}}, 
\end{equation}
where the index $\alpha$ corresponds to the sub-binning of the bin
$\mathcal{D}_i$. 

\begin{figure}
  \epsfig{figure=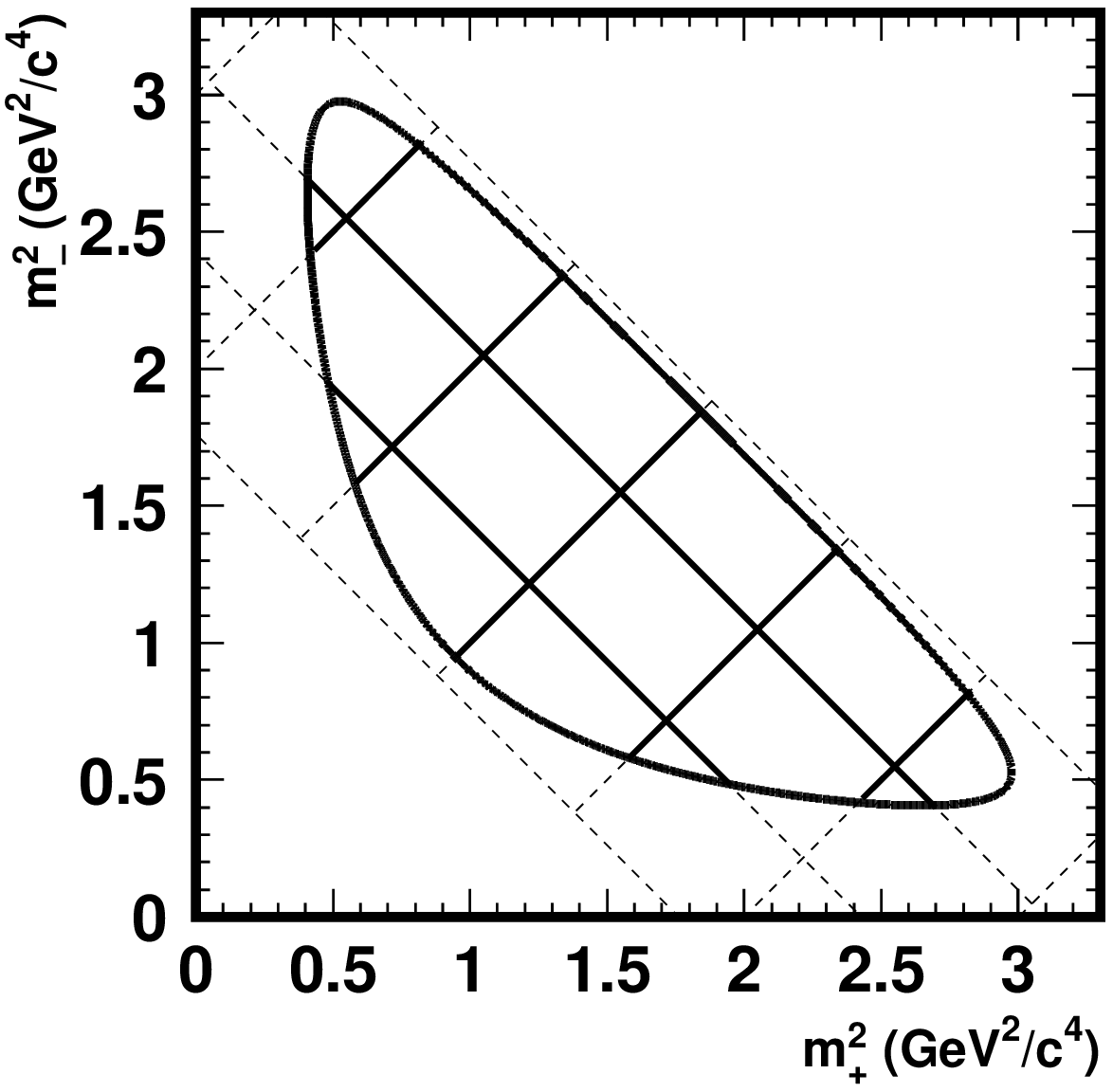,width=0.3\textwidth}
  \hfill
  \epsfig{figure=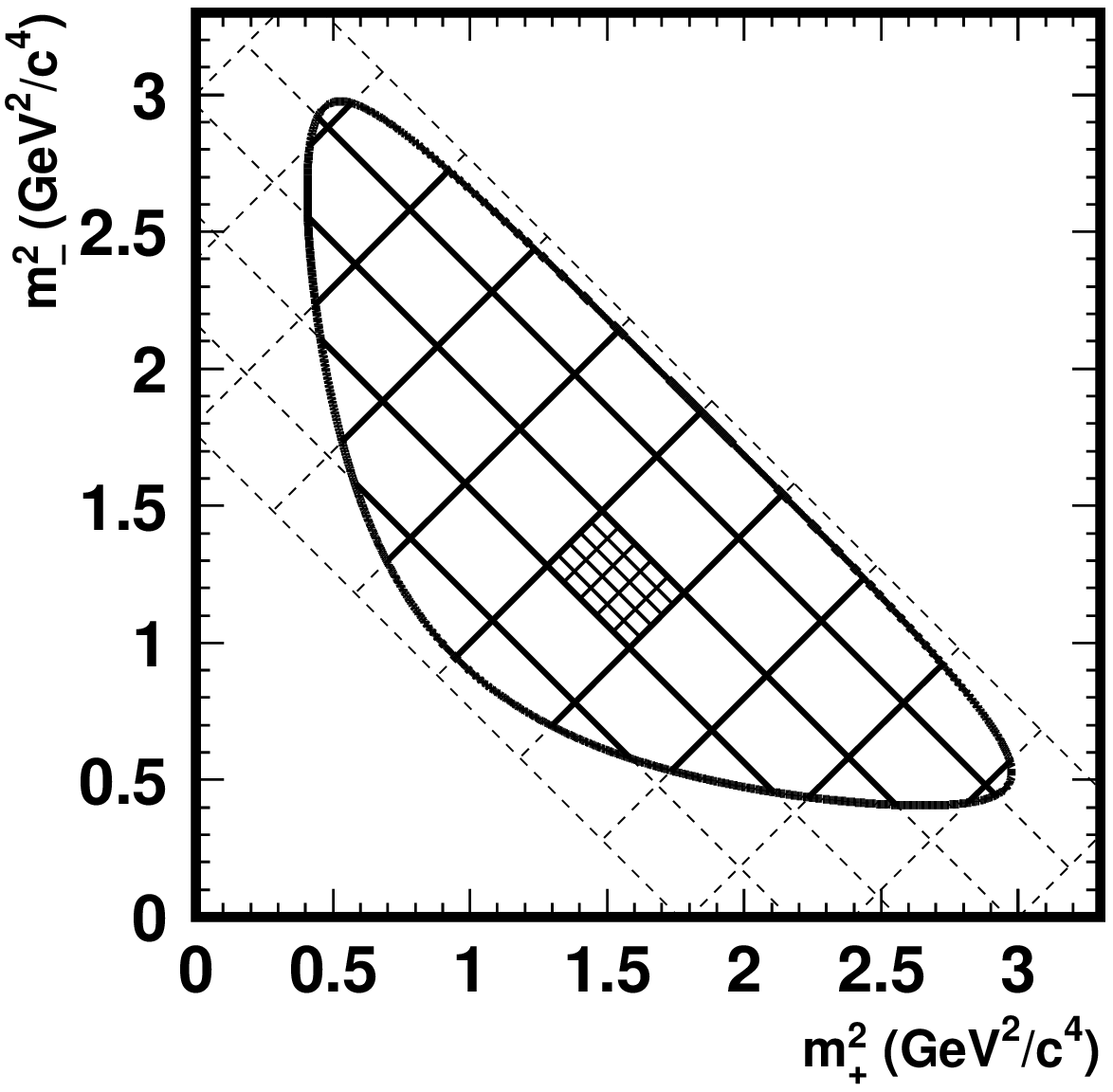,width=0.3\textwidth}
  \hfill
  \epsfig{figure=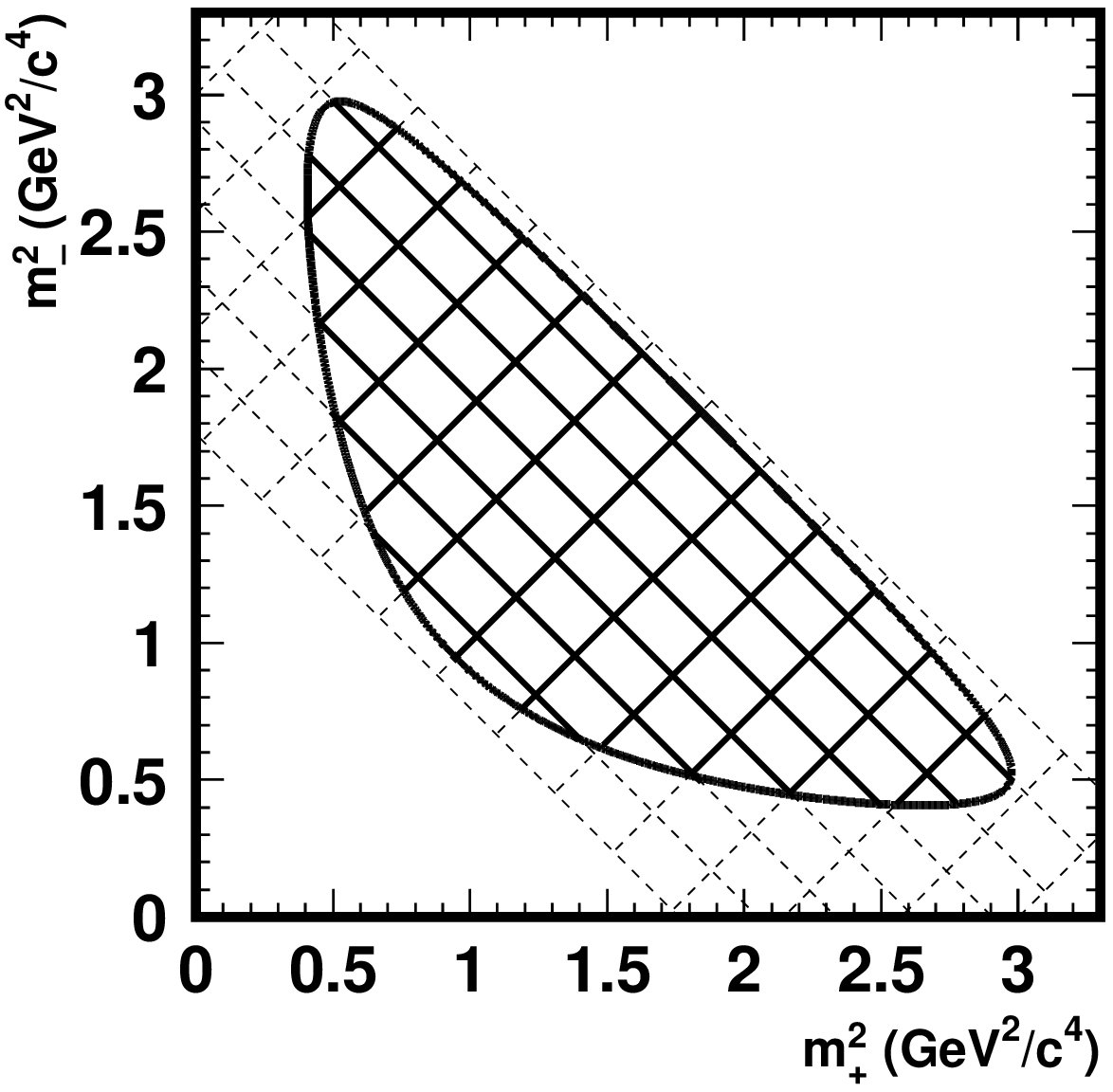,width=0.3\textwidth}
  \caption{Binning of the \dkpp\ Dalitz plot. 3x3 (left), 5x5 (center)
           and 7x7 binning (right). 5x5 sub-binning of one of the bins 
           of the central plot is shown; the sub-binning is used for 
           $D_{CP}$ Dalitz plot. }
  \label{plot}
\end{figure}

A toy MC study has been performed to obtain the accuracy of $\phi_3$
measurement as a function of $B$ and $D_{CP}$ decay statistics, and to 
investigate the influence of the finite bin size on the measurement result. 
The Dalitz plot binning used for our MC study is
shown in Fig.~\ref{plot}. 
%The phase space 
%$\Delta m^2=m^2_+-m^2_-\in (-3\mbox{ GeV}^2/c^2, 3\mbox{ GeV}^2/c^2)$
%and $m^2_{\pi\pi}=M^2_D-2M^2_{\pi}-M^2_{K}-m^2_+-m^2_-\in 
%(0, 2\mbox{ GeV}^2/c^2)$ was uniformly divided into $10\times 5$ rectangular
%bins. 38 of 50 bins intersect the kinematically allowed phase space 
%for \dkpp\ decay, therefore, $\mathcal{N}=19$. 
The phase space 
$\Delta m^2=m^2_+-m^2_-\in (-3\mbox{ GeV}^2/c^4, 3\mbox{ GeV}^2/c^4)$
and $m^2_{\pi\pi}=M^2_D-2M^2_{\pi}-M^2_{K}-m^2_+-m^2_-\in 
(0, 2\mbox{ GeV}^2/c^4)$ was uniformly divided into rectangular
bins, symmetrically over the exchange of the sign of the final state pions. 
The bin size should certainly affect the statistical precision 
of the measurement. Since the structure of the \dkpp\ decay matrix element 
is quite complex, the variations of the event density due to $CP$ violation
change rapidly across the phase space, and, as the bin size becomes larger, 
the average effect of $CP$ violation on each bin is reduced. 
Three different binnings were studied; depending on the number of bins in 
half of the phase space they are referred to as 3x3 (the corresponding 
number of bins which enter the half of the kinematically allowed phase 
space is $\mathcal{N}=8$), 5x5 ($\mathcal{N}=19$) and 7x7 ($\mathcal{N}=32$)
binnings. 

The parameters $r_B$ and $\delta_B$
were taken to be $0.2$ and $180^{\circ}$, respectively; $\phi_3$ was 
scanned in a range from $0^{\circ}$ to $180^{\circ}$. For the $D^0$
decay model, the recent Belle measurement was used \cite{phi3_belle_3}. 
The efficiency variations over the phase space and background 
contribution were not taken into account in the simulation. 
The statistical precision of the method certainly depends on the 
actual $D^0$ model, therefore, the real experimental precision can 
somewhat differ from our estimate. The error is also inversely 
proportional to the magnitude $r_B$ of the opposite $D$ flavor contribution. 
The estimates of this value vary in the range 0.1--0.2, the current
measurements of $r_B$ \cite{phi3_belle_4,phi3_babar_2} agree with 
these numbers, although the error is still large. If 
the actual value of $r_B$ differs from our expectation, the results of our
study should be scaled accordingly. 

\begin{figure}
  \epsfig{figure=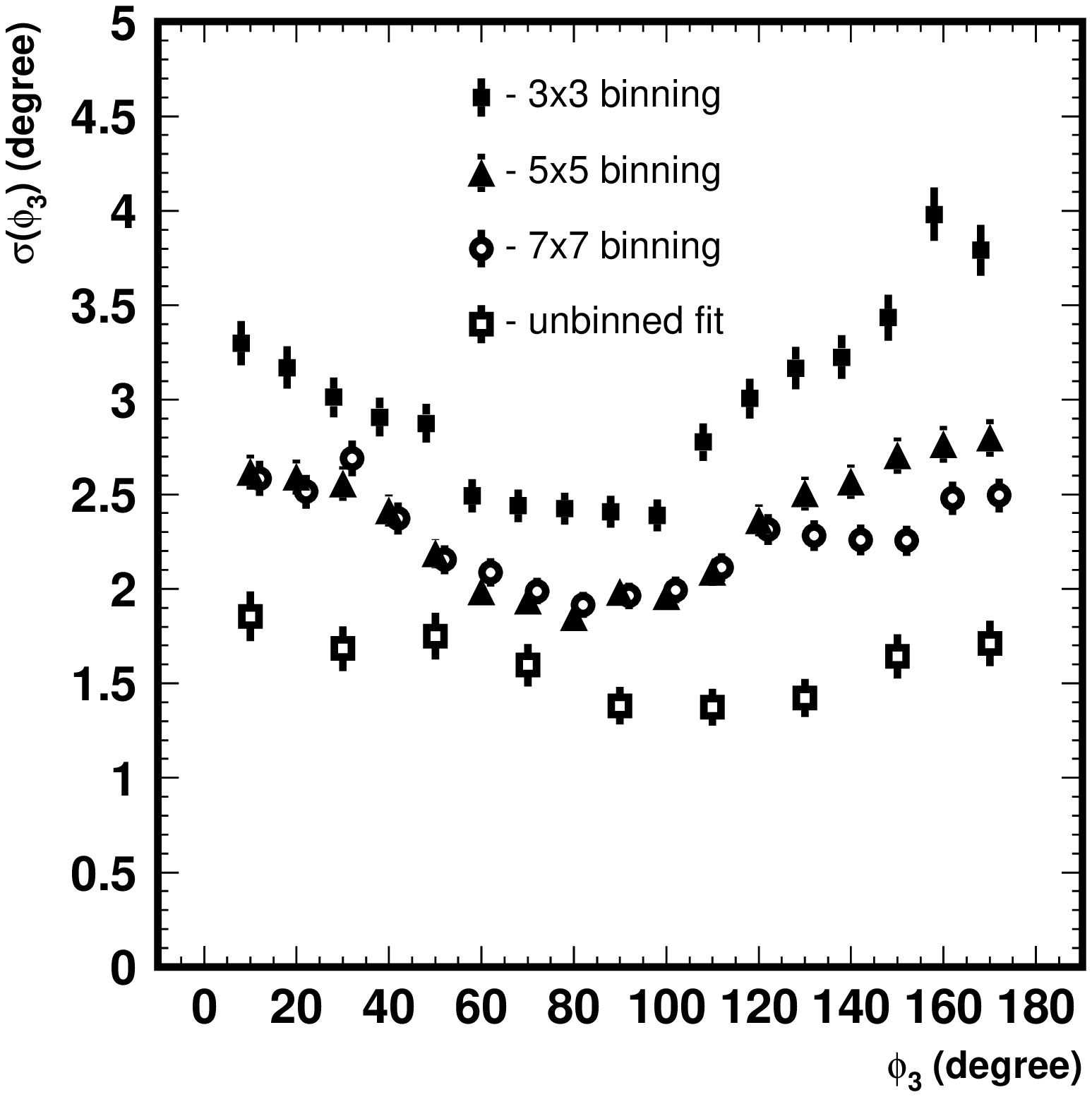,width=0.48\textwidth}
  \hfill
  \epsfig{figure=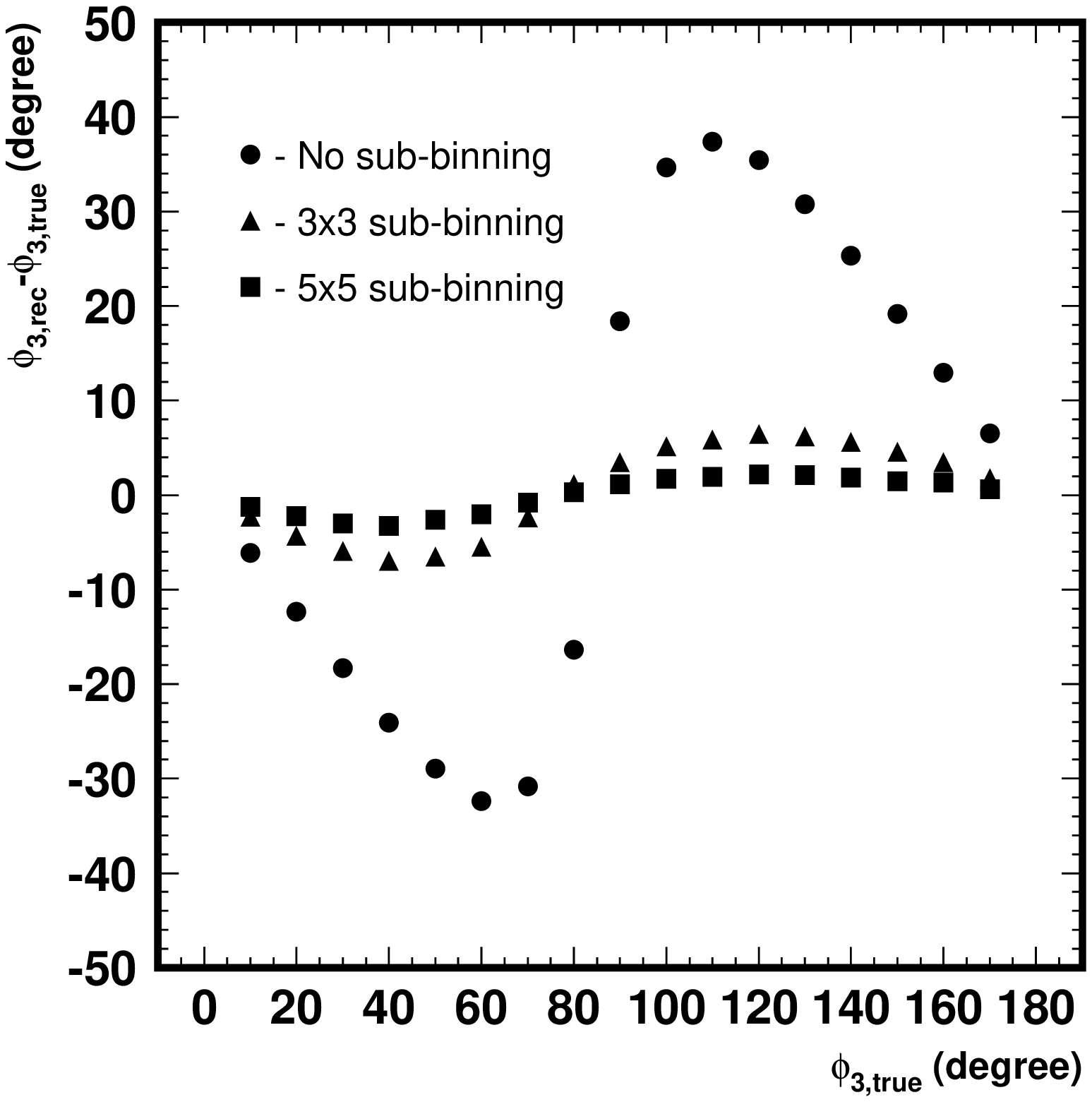,width=0.48\textwidth}

  \parbox[t]{0.47\textwidth}{
  \caption{$\phi_3$ error as a function of $\phi_3$, for different 
           Dalitz plot binnings. Results of the toy MC study with 
           $2\times 10^4$ \bdk\ decays, $4\times 10^5$ 
           $D_{CP}$ decays, $r_B=0.2$, $\delta_B=180^{\circ}$. }
  \label{sigcomp}
  }
  \hfill
  \parbox[t]{0.47\textwidth}{
  \caption{$\phi_3$ bias due to finite size of the bin as a function of 
           $\phi_3$, for the cases of no 
           sub-binning (circles), 3x3 sub-binning (triangles), 
           5x5 sub-binning (squares). }
  \label{binsize}
  }
\end{figure}

The statistical error of the $\phi_3$ measurement as a function of $\phi_3$
for the different binnings is shown in Fig.~\ref{sigcomp}.
For each $\phi_3$ value, 200 pseudo-experiments with 
the statistics of $10^4$ \bdk\ decays were generated. The $\phi_3$ error was 
obtained from the spread of the reconstructed values. To keep the 
contribution of finite $D_{CP}$ statistics small, we used the large 
number of $D_{CP}$ decays ($2\times 10^5$ decays of each $CP$-parity).
%The statistics of $10^4$ \bdk\ decays 
%of each flavor, and $2\times 10^5$ $D_{CP}$ decays of each $CP$-parity were 
%used. 
The $\phi_3$ accuracy depends only weakly on the values of $\phi_3$
or $\delta_B$ since both $c_i$ and $s_i$ are 
obtained from the same $D_{CP}$ statistics and have comparable accuracies. 
There is no significant difference in the statistical accuracy for 5x5 and
7x7 binnings, while in the case of 3x3 binning the accuracy is worse 
by 30-40\%. We have used the 5x5 binning for all further studies (excluding
the study of the approach without $D_{CP}$ decays, see 
Section~\ref{section_no_dcp}). The precision of $\phi_3$ extracted 
from a model-dependent unbinned Dalitz plot fit with the same input parameters 
is also shown for comparison. The unbinned fit is the optimal technique from 
the statistical point of view and represents the best limit of the statistical 
precision. The resolution in that case is approximately 30\% better than 
for the case of binned approach. The optimal choice of binning can 
possibly improve the accuracy of the binned technique. We leave that 
question for the future studies. 

The results of the toy MC study of $\phi_3$ bias due to finite $D$-bin 
size are shown in Fig.~\ref{binsize}. The angle $\phi_3$
was scanned from 0 to 180$^{\circ}$, other parameters were taken to 
be $r_B=0.2$, $\delta_B=180^{\circ}$.  Points show the $\phi_3$ bias for no sub-binning, 3x3 sub-binning 
and 5x5 sub-binning (shown in Fig.~\ref{plot} for one of the bins). 
For our further study, we have used 5x5 sub-binning, for which 
the bias does not exceed 2$^{\circ}$. In a real experiment the bias
can be corrected, {\it e.g.} based on the quasi two-body model. 
Since the correction is comparatively small, the uncertainty introduced
by a model-based correction should be negligible. 

\begin{figure}
  \epsfig{figure=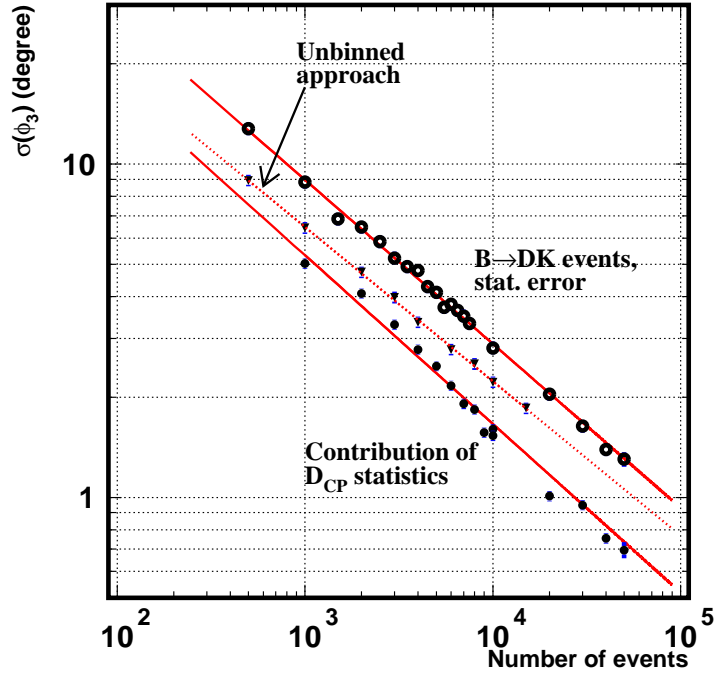,width=0.55\textwidth}
  \caption{Statistical error of $\phi_3$ as a function of number of 
           reconstructed \bdk\ decays (open circles) and $D_{CP}$ decays (dots). 
           Results of toy MC study with $r_B=0.2$, 
           $\phi_3=70^{\circ}$, $\delta_B=180^{\circ}$. 
           $\phi_3$ error from model-dependent unbinned Dalitz plot fit 
           with the same input parameters is also shown (dotted line). }
  \label{cpstat}
\end{figure}

The results of the study of $\phi_3$ sensitivity dependence
on the number of reconstructed \bdk\ and $D_{CP}$ decays
are shown in Fig.~\ref{cpstat}. The graph
shows the $\phi_3$ error for $r_B=0.2$, $\phi_3=70^{\circ}$, $\delta_B=180^{\circ}$.
For different values of $r_B$, the error scales inversely proportional to $r_B$.
The weak dependence of the statistical error on the phases $\phi_3$
and $\delta_B$ should also be noted, therefore, the actual $\phi_3$ 
error can differ by 30--50\% depending on the true values of these phases. 
The dependences on both $B$ and $D_{CP}$ statistics show a square root scaling. 
For comparison, the statistical error of $\phi_3$ extracted by 
model-dependent unbinned Dalitz plot fit with the same input parameters 
is shown with the dotted line. 

%The statistical error of the current Belle $\phi_3$ 
%measurement, obtained with unbinned Dalitz plot analysis
%(filled triangle in Fig.~\ref{cpstat}), agrees well with the $B$ decay 
%statistics dependence. This indicates that the binning procedure does not 
%affect significantly the statistical accuracy of the measurement. 

According to our estimations, the proposed super-$B$ factory with its 
design integrated luminosity of 50~ab$^{-1}$, would allow a measurement of 
$\phi_3$ with accuracy below 2$^{\circ}$. To keep the contribution to the 
statistical error from the finite sample of $D_{CP}$ decays below that 
level, one needs about $10^4$ tagged $D_{CP}$ decays, corresponding to
approximately 10~fb$^{-1}$ of integrated luminosity at $\psi(3770)$ peak. 
In the following sections we consider possible alternative approaches 
of the binned analysis and compare their precision with the approach 
described above, which we have found to be the most optimal.

\section{Approach without $s_i$ constraint}

\begin{figure}
  \epsfig{figure=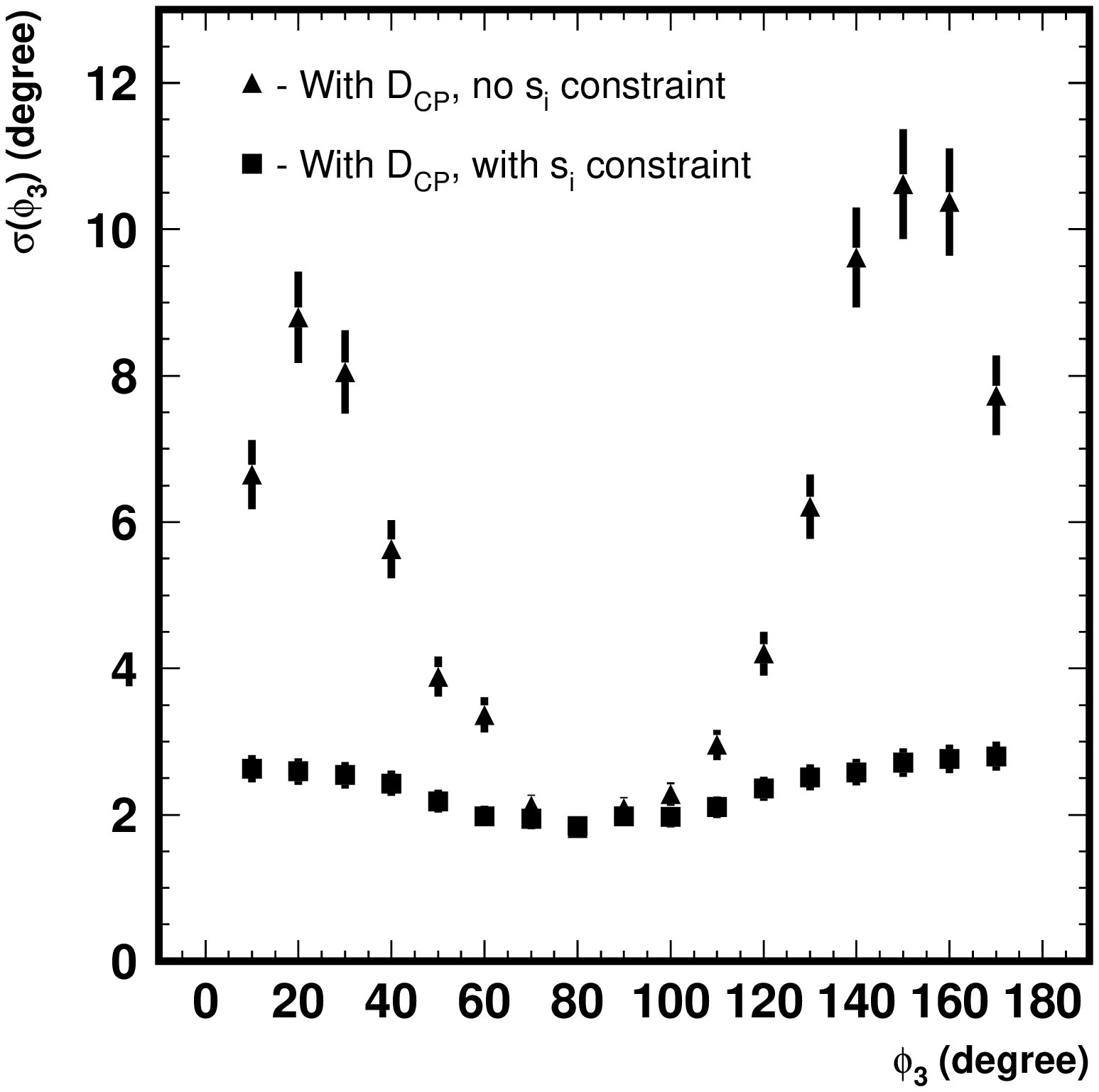,width=0.47\textwidth}
  \hfill
  \epsfig{figure=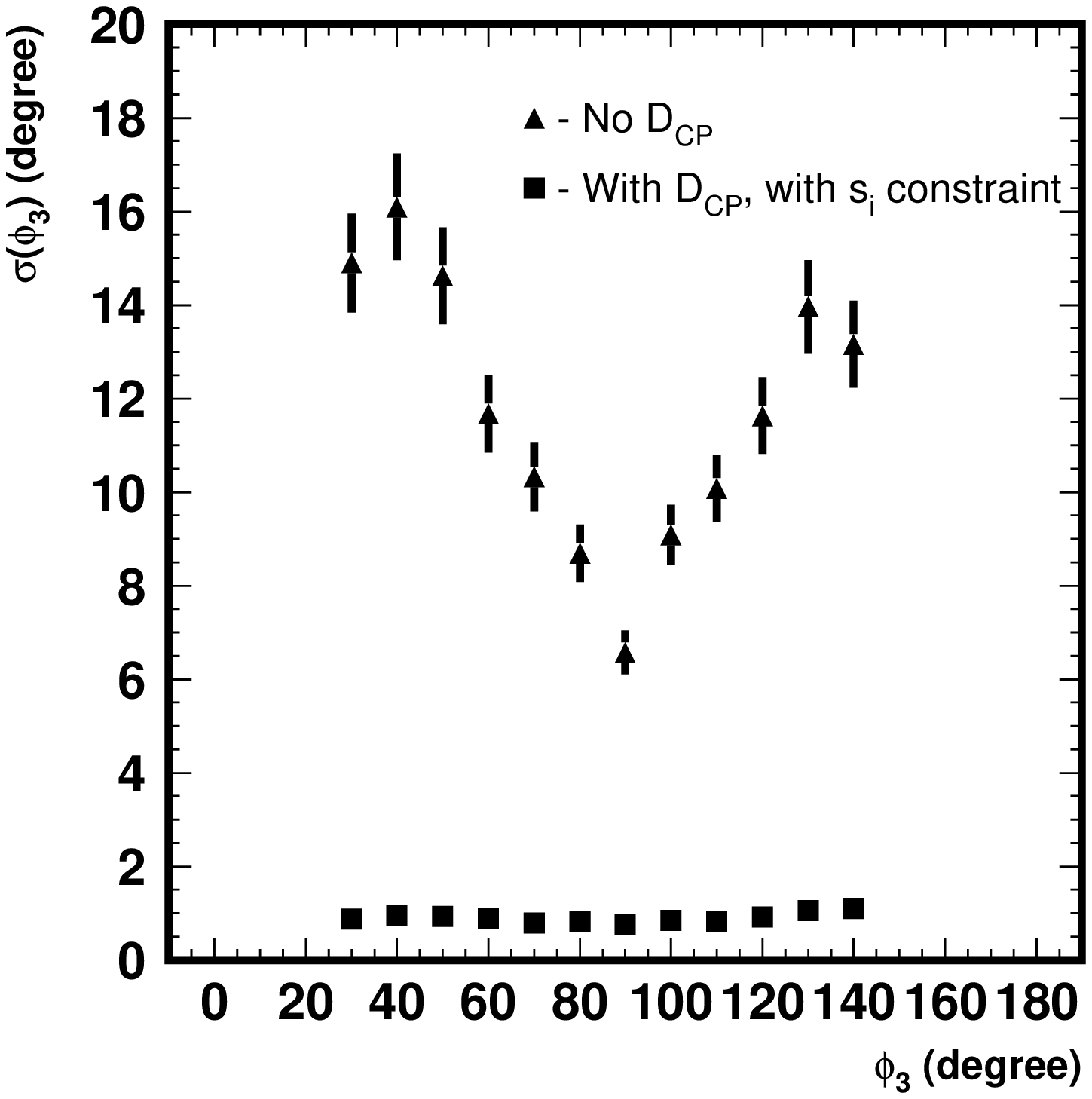,width=0.47\textwidth}
  \parbox[t]{0.47\textwidth}{
  \caption{$\phi_3$ error as a function of $\phi_3$ for relaxed $s_i$ 
    (triangles); and $s_i$ constrained from $c_i$ (squares). 
    Results of the toy MC study with $2\times 10^4$ \bdk\ decays, $4\times 10^5$ 
    $D_{CP}$ decays, $r_B=0.2$, $\delta_B=180^{\circ}$, 5x5 Dalitz plot binning. }
  \label{tagged_dcp}
  }
  \hfill
  \parbox[t]{0.47\textwidth}{
  \caption{$\phi_3$ error as a 
  function of $\phi_3$ for the model-independent approach without $D_{CP}$
  decays (triangles) and with $D_{CP}$ decays and $s_i$ constraint 
  (squares). Results of the toy MC study with $2\times 10^5$ \bdk\ decays, 
  $r_B=0.2$, $\delta_B=180^{\circ}$, 3x3 Dalitz plot binning. }
  \label{nocp_sig}
  }
  
\end{figure}

Constraining $s_i$ from $c_i$ introduces a slight model dependence to the 
measurement result, since the bin size becomes dependent on the width 
of the Dalitz plot structures of $D^0$ decay. In principle, the system 
of equations is solvable even without this constraint. 
In this case, 
there are $8\mathcal{N}$ equations (\ref{bnum}, \ref{cpnum}) for  
$2\mathcal{N}+6$ unknowns ($c_i$, $s_i$, $r_B$, $\phi_3$, $\delta_B$, $h_B$, 
$h_{CP\pm}$). 
Coefficients $c_i$ are still directly obtained from the $D_{CP}$ decays, 
while $s_i$ are mainly constrained by the $B$ decays. 
Therefore, we expect the accuracy of the 
$\cos(\pm\phi_3+\delta_B)$ determination to be higher than that for 
$\sin(\pm\phi_3+\delta_B)$. In other words, the accuracy of $\phi_3$
determination should strongly depend on the values of $\phi_3$ and $\delta_B$. 
The $\phi_3$ error as a function of $\phi_3$
for $r_B=0.2$ and $\delta_B=180^{\circ}$ is shown in Fig.~\ref{tagged_dcp}
(triangles). This plot was obtained by toy MC
procedure with $10^4$ \bdk\ decays of each flavor, and $2\times 10^5$ 
$D_{CP}$ decays of each $CP$-parity. For comparison, the $\phi_3$ error is 
shown obtained with the approach where $s_i$ is fixed from $c_i$. 
As follows from Fig.~\ref{tagged_dcp}, the precision of the approach 
under study is comparable with the optimal strategy precision only in the 
narrow range of $\phi_3$ values. The position of this region
depends on the value of the strong phase $\delta_B$, therefore, the 
precision of the $\phi_3$ determination in this approach can not be 
guaranteed.

\section{Approach without tagged \boldmath{$D_{CP}$}}

\label{section_no_dcp}

Even if the $D_{CP}$ decays are not available, it is in principle possible 
to extract $\phi_3$ in a model-independent way using only \bdk\ and
flavor-tagged $D$ decays. 
For $\mathcal{N}$ pairs of bins, there are $4\mathcal{N}$ equations (\ref{bnum})
($2\mathcal{N}$ for each flavor of $B$), and $2\mathcal{N}+4$ unknowns
($c_i$, $s_i$, $r_B$, $\phi_3$, $\delta_B$, $h_B$). Therefore, the system is solvable 
for $\mathcal{N}\geq 2$. Technically (if the number of equations is greater than
the number of unknowns), it can be done by minimizing the negative
logarithmic likelihood (Eq.~\ref{lhsum}, \ref{lh}) with $c_i$ and $s_i$
coefficients taken as free parameters. 

Practically, however, this approach offers too low sensitivity for any 
reasonable $B$ decay statistics. A toy MC study has been performed
with $10^5$ \bdk\ decays of each flavor. 
%, and Dalitz plot binning was
%the same as in the previous case (see Fig.~\ref{plot}). 
The parameters $r_B$ and $\delta_B$
were taken to be $0.2$ and $180^{\circ}$, respectively; $\phi_3$ was 
scanned in a range from $30^{\circ}$ to $140^{\circ}$. 
The results of the toy MC study are presented in Fig.~\ref{nocp_sig} for 
3x3 binning. Statistical precision with $D_{CP}$ data and $s_i$ constraint
is also shown for comparison. It can be seen from the plots, that the 
accuracy of the $\phi_3$ determination is quite poor compared to the method 
which makes use of $D_{CP}$ decays. Reaching the accuracy of $2^{\circ}$
would require about $5\times 10^6$ reconstructed \bdk\ decays, which 
corresponds to the currently unrealistic luminosity integral of $5000$ ab$^{-1}$ 
at $B$-factory.

Taking finer 5x5 binning could slightly improve the sensitivity, however, 
due to large number of free parameters, the fit converges poorly even with 
a number of $B$ decays as large as $10^5$. 
Variations of the initial parameters in the fit can lead to significant 
changes of the fit result. Binning finer than 
5x5 has not been investigated due to an internal limitation on the maximum 
number of free parameters in the software used for minimization. 

%\begin{figure}
%  \epsfig{figure=nocp_sig.eps,width=\textwidth}
%  \caption{(left) residual $\phi_3$ distribution for 
%  $\phi_3=70^{\circ}$ (open histogram) and $\phi_3=140^{\circ}$
%  (hatched histogram) and (right) $\phi_3$ RMS as a 
%  function of $\phi_3$ for the model-independent approach without $D_{CP}$
%  decays (squares) and with $D_{CP}$ decays and $s_i$ constraint 
%  (triangles). Results of the toy MC study with $10^5$ \bdk\ decays, 
%  $r=0.2$, $\delta=180^{\circ}$. }
%  \label{nocp_sig}
%\end{figure}

\section{Conclusion}

We have performed a toy Monte-Carlo study of a model-independent
way to measure the angle $\phi_3$ of the unitarity triangle, 
suggested by Giri {\it et al.} \cite{phi3_modind}. The method 
involves \bdk\ decays with neutral $D$ decaying to $K^0_S\pi^+\pi^-$ final state, 
together with the decays of $D$ mesons in $CP$ eigenstate to the same 
final state. Different approaches to the extraction of $\phi_3$
are considered. Although technically the model-independent
measurement of $\phi_3$ can be performed without decays of $D$
in $CP$ eigenstate, the statistics of \bdk\ decays required to obtain 
a reasonable $\phi_3$ accuracy in this approach is unrealistically high. 

We have shown that for the optimal strategy (which involves constraining
$s_i$ coefficients from $c_i$), the accuracy of the 
model-independent binned approach is only 30\% worse compared to the unbinned 
technique. Possibly the choice of the optimal binning ({\it e.g.} using
non-rectangular bins) can improve the accuracy of the model-independent 
measurement. The statistical error of $\phi_3$ 
does not depend significantly on the $\phi_3$ and strong phase values. 

According to our estimations, the proposed super-$B$ factory \cite{superb}
with its design integrated luminosity of 50~ab$^{-1}$, would allow a 
measurement of $\phi_3$ with statistical accuracy below 2$^{\circ}$. 
To keep the contribution to the 
statistical error from the finite sample of $D_{CP}$ decays below that 
level, around $10^4$ tagged $D_{CP}$ decays are needed, corresponding to
approximately 10~fb$^{-1}$ of integrated luminosity collected at 
$\psi(3770)$ peak. This task can be accomplished by CLEO-c experiment 
at CESR collider \cite{cleoc} and by upgraded BESIII/BEPCII complex at 
Beijing with design luminosity of $10^{33}$ cm$^{-2}$s$^{-1}$ \cite{bes}. 

The systematic errors due to variations of the detection efficiency or
background density over the Dalitz plot and other related factors were 
not considered in this study. 
%Other contributions to the systematic error, such as background contamination
%$or detection efficiency variations, were not considered in our study. 
They will certainly depend on 
the specific detector parameters, but the current experience with
model-dependent Dalitz analyses at $B$ factories allows us to believe that 
these contributions will also decrease with increasing statistics. 

\section{Acknowledgments}

We would like to thank Tim Gershon and David Asner for helpful discussions
and suggestions on improving this paper. We are grateful to the Belle 
collaboration for the interest expressed to our study. 

This work is supported by the Federal Agency for Science and Innovations 
of the Russian Federation under contract 02.434.11.7075.

\end{document}